\documentclass[%
 reprint,
 amsmath,amssymb,
 aps,
 prl,
longbibliography
]{revtex4-2}
\usepackage{lipsum}
\usepackage{xcolor}
\usepackage{soul}
\usepackage{graphicx}
\usepackage{dcolumn}
\usepackage{bm}
\usepackage[colorlinks = true,
            linkcolor = blue,
            urlcolor  = blue,
            citecolor = blue,
            anchorcolor = red]{hyperref}

\begin{document}


\title{Experimental realization of supergrowing fields}%

\author{Sethuraj K. R.$^{1,2}$}
\email{skarimpa@ur.rochester.edu}
\author{Tathagata Karmakar$^{3,2,4}$}
\email{tkarmaka@ur.rochester.edu}
\author{S. A. Wadood$^{1,2,5}$}
\author{Andrew N. Jordan$^{4,3,2}$}
\author{A.~Nick Vamivakas$^{1,2,3,6}$}
\email{nick.vamivakas@rochester.edu}
\affiliation{$^1$The Institute of Optics, University of Rochester, Rochester, NY 14627, USA}
\affiliation{$^2$Center for Coherence and Quantum Optics, University of Rochester, Rochester, NY 14627, USA}
\affiliation{$^3$Department of Physics and Astronomy, University of Rochester, Rochester, NY 14627, USA}
\affiliation{$^4$Institute for Quantum Studies, Chapman University, Orange, CA 92866, USA}%
\affiliation{$^5$Currently with Department of Electrical Engineering, Princeton University, NJ, 08544, USA}
\affiliation{$^6$Materials Science, University of Rochester, Rochester, NY 14627, USA}




\date{\today}

\begin{abstract}

Supergrowth refers to the local amplitude growth rate of a signal being faster than its fastest Fourier mode.
In contrast, superoscillation pertains to the variation of the phase. 
Compared to the latter, supergrowth can have exponentially higher intensities and promises improvement over superoscillation-based superresolution imaging.
Here, we demonstrate the experimental synthesis of controlled supergrowing fields with a maximum growth rate of $\sim 19.1$ times the system-bandlimit.
Our work is an essential step toward realizing supergrowth-based far-field superresolution imaging. 


\end{abstract}

\maketitle

Superresolution in optical imaging refers to approaches that can boost spatial resolution beyond the diffraction limit of light. The diffraction limit defines the smallest feature size that can be resolved in a standard optical imaging system and is determined by the light’s wavelength and the optical system's numerical aperture (NA) \cite{goodmanIntroductionFourierOptics2017}. One way to resolve subwavelength features 
in far-field imaging is by using superoscillatory optical spots, a phenomenon where complex fields can locally oscillate at a  rate greater than their cut-off spatial frequency \cite{Aharonov-preprint,chen_superoscillation_2019, Gbur+2019+SO_SR_review, Berry2006May,berry2019roadmap}. 
Nonetheless, superoscillations have the inherent disadvantage of very small intensity combined with substantial side lobes, leading to poor imaging quality. Both numerical optimization schemes \cite{rogersOptimisingSuperoscillatorySpots2018} and sophisticated optical setups \cite{hu_optical_2021, Kozawa2018Feb,Dong:17} have been investigated for the mitigation of sidelobe intensity. However,  supergrowth \cite{jordanSuperphenomenaArbitraryQuantum2022}, a recently introduced physical concept, offers a promising route to solve this issue.

In supergrowing fields, the local amplitude growth rate of a complex field is higher than the highest spatial frequency in its Fourier spectrum, thereby also providing access to subwavelength features \cite{jordanSuperresolutionUsingSupergrowth2020}. This concept has parallels to near-field microscopy with evanescent waves \cite{pohl_optical_1984,yang_super-resolution_2014}. 
Supergrowing optical field spots can contain exponentially more intensity compared to superoscillating regions and have been shown theoretically to be able to image subwavelength objects \cite{karmakar2023supergrowth}. 
Moreover, theoretical frameworks for systemic generation of both supergrowing/superoscillating fields have been studied \cite{karmakar2023superoscillation}. 
Currently,  supergrowth exists as a purely theoretical construct and has not been realized experimentally. 

In this work, we achieve a critical step toward the goal of superresolution using supergrowth, i.e.,~the synthesis and characterization of supergrowing fields in the laboratory. 
To accomplish this objective, we elucidate the concept of supergrowth in a diffraction-limited system and identify the critical constraints that can limit supergrowth in an experiment. We also furnish comprehensive methods and tools to  quantify supergrowth in any experimental setup effectively.
The experimental demonstration lays the foundation for  generating supergrowing fields that  can be utilized for superresolution imaging.


\emph{Supergrowing strength.}---
The local growth rate  $\kappa(x)$ [local wave number $k(x)$] of a bandlimited complex-field $f(x)$ is defined as $\kappa(x)+ik(x)=\partial_x \log \big[f(x) \big]$  \cite{jordanSuperresolutionUsingSupergrowth2020}. 
The field has supergrowth (superoscillation) at $x=x_0$ when $|\kappa(x_0)|\geq k^f_{\max}$ ( $|k(x_0)|\geq k^f_{\max}$), where $k^f_{\max}$ denotes the highest wavenumber of the field. 
To characterize the amount of supergrowth, we define the supergrowing strength $\Gamma(x) =\big|\frac{\kappa(x)}{k^f_{\max}}\big|$; $f(x)$ is supergrowing at $x=x_0$ when  $\Gamma(x_0)\geq 1$.

The intensity associated with field $f(x)$ is $I(x)=|f(x)|^2$, and has  a bandlimit of $2k^f_{max}$. The local growth rate of the intensity $\kappa_I(x)$ is twice that of the original field. The supergrowing strength $\Gamma(x)$ can be obtained from the intensity as 
\begin{equation}
    \Gamma(x)=\Big|\frac{\kappa_I(x)}{2k^{f}_{\max}}\Big|.
    \label{Gamma_s_intensity_eq}
\end{equation}

Equation.~\eqref{Gamma_s_intensity_eq} has significant implications in the experiment, as we need only transverse intensity information, which can be obtained using a CCD camera. For field-based experimental characterization of growth rate, one needs full-field reconstruction involving more complicated interferometry.

\begin{figure}[h]
\includegraphics[width=8.5cm]{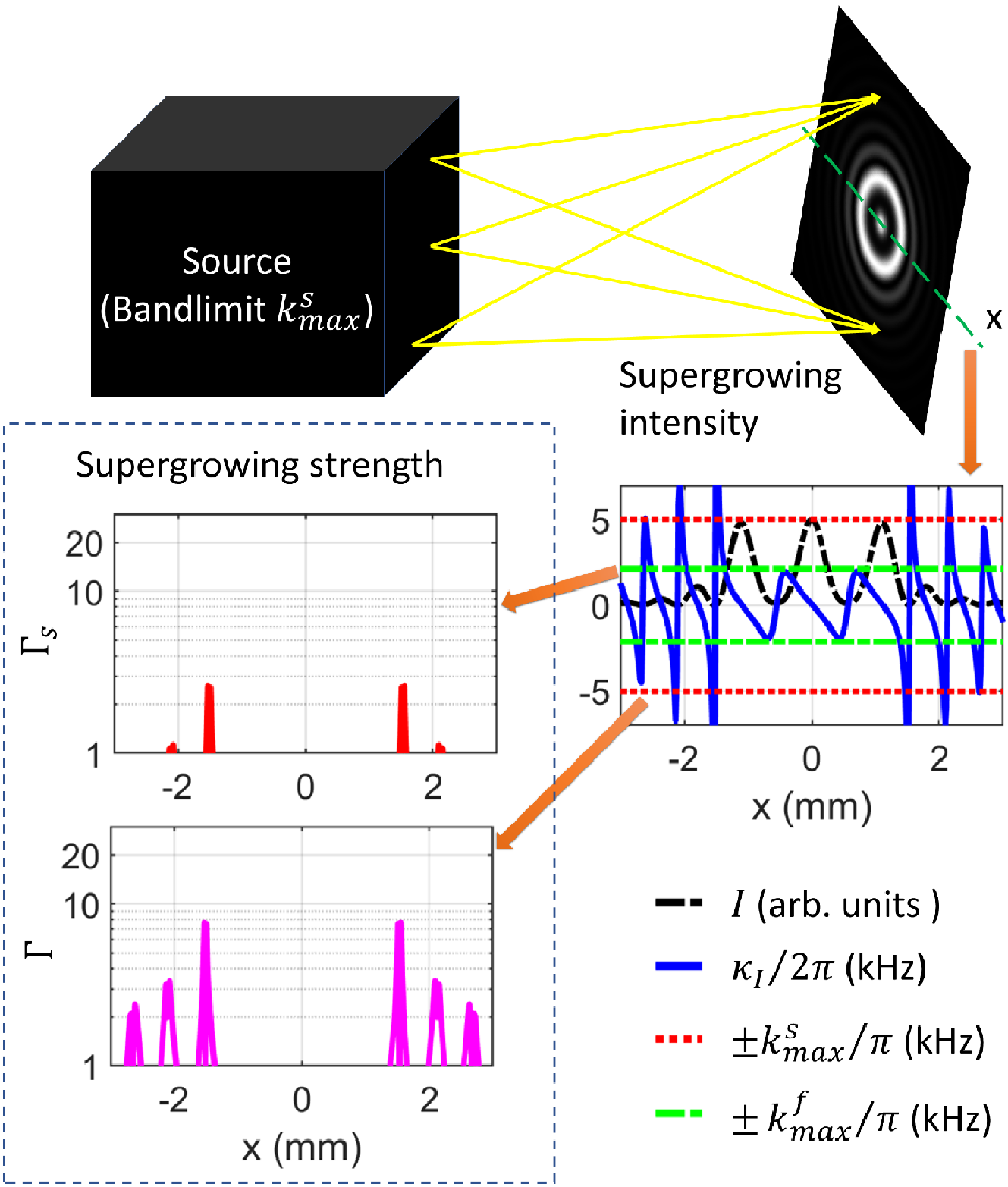}
\caption{\label{fig1} Conceptual illustration of the dependence of supergrowth on field and system bandlimits. 
The optical source (shown as a black box) of bandlimit $k^s_{\max}$ synthesizes a supergrowing complex field (of bandlimit $k^f_{\max}$) at the image plane. 
Horizontal line-cut of intensity $I(x)$, and supergrowing strength parameters $\Gamma(x)$  from Eq.~\eqref{Gamma_s_intensity_eq} and $\Gamma_s(x)$ from Eq.~\eqref{Gamma_s_intensity_eq_sys} are plotted for a sample function.
Since $k^s_{max} \geq k^f_{max}$ in any diffraction-limited optical system, the condition $\Gamma_s\geq 1$ is more restrictive in comparison to  $\Gamma\geq 1$. This fact is manifested in the plot since a smaller region is identified by $\Gamma_s\geq 1$.  
} 
\end{figure}

\emph{Supergrowth in a diffraction-limited optical system.}---The wavelength of illumination $\lambda$ and an optical system's NA define its diffraction-limited response and maximum spatial frequency ($k^s_{\max}\approx\tfrac{2\pi}{\lambda}\mathrm{NA}$) that can pass through the system \cite{goodmanIntroductionFourierOptics2017}.  
The bandlimit of the optical system might be different from the field bandlimit $k^f_{\max}$, although the latter must be upper bounded by the former (i.e.~$k^f_{\max}\leq k^s_{\max}$). 
We note that superresolution imaging using superoscillation/supergrowth is only possible when the local wavenumber/growth rate of the illumination field is higher than the bandlimit  of the system $k^s_{max}$. For $k^f_{max}\leq |k|,|\kappa| < k^s_{max}$,   superresolution  cannot be achieved.

For supergrowth in experiments, the appropriate definition is $|\kappa(x)|\geq k^s_{\max}$, so that the benefit of supergrowth can outperform the diffraction-limited performance of an optical system. The supergrowing strength of an optical field in an experiment is 
\begin{equation}
    \Gamma_s(x)=\Big|\frac{\kappa_I(x)}{2k^{s}_{\max}}\Big|,
    \label{Gamma_s_intensity_eq_sys}
\end{equation}
where the subscript $s$ in $\Gamma_s$ indicates that it is defined with respect to the bandlimit of the system. According to this definition, the field $f(x)$ is supergrowing when $\Gamma_s \geq 1$. The optical fields realized in our experiment are supergrowing with respect to both definitions. For convenience, we define the bandlimit ratio $b=(k^{f}_{\max})/(k^{s}_{\max})$, such that $\Gamma_s=b~\Gamma$. 

Figure~\ref{fig1} provides an example of an optical field with   different $\Gamma_s(x)$ and $\Gamma(x)$. One can observe that at some points  $x=x_0$, the optical field is supergrowing  in a mathematical sense ($\Gamma \geq 1$), but not supergrowing ($\Gamma_s < 1$) based on the definition we use. From here onwards, we stick to the definition of supergrowth strength as $\Gamma_s$.

\begin{figure}[b]
\includegraphics[width=8.5cm]{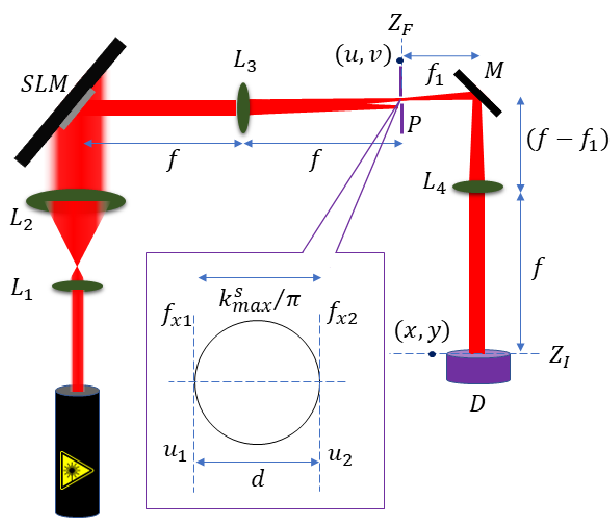}
\caption{\label{fig2} Optical setup schematic.
Notations: $L$, lens; $SLM$, Spatial light modulator; $f$, focal length of lens; $Z_F$; transverse back focal plane of $L_3$; $P$, precision pinhole; $M$, mirror; $D$, CCD camera; $Z_I$, image plane; $(u,v)$, Cartesian coordinates representing any point on the plane $Z_F$; $(x,y)$, Cartesian coordinates representing any point on the plane $Z_I$; $d$, diameter of $P$; and $f_{x1}$ and $f_{x2}$, are horizontal ($x$) Fourier coordinate of $Z_I$. 
The inset shows an expanded view of the transverse cross-section of $P$.}
\end{figure}

\begin{figure*}[t]
\centering
\includegraphics[width=\textwidth]{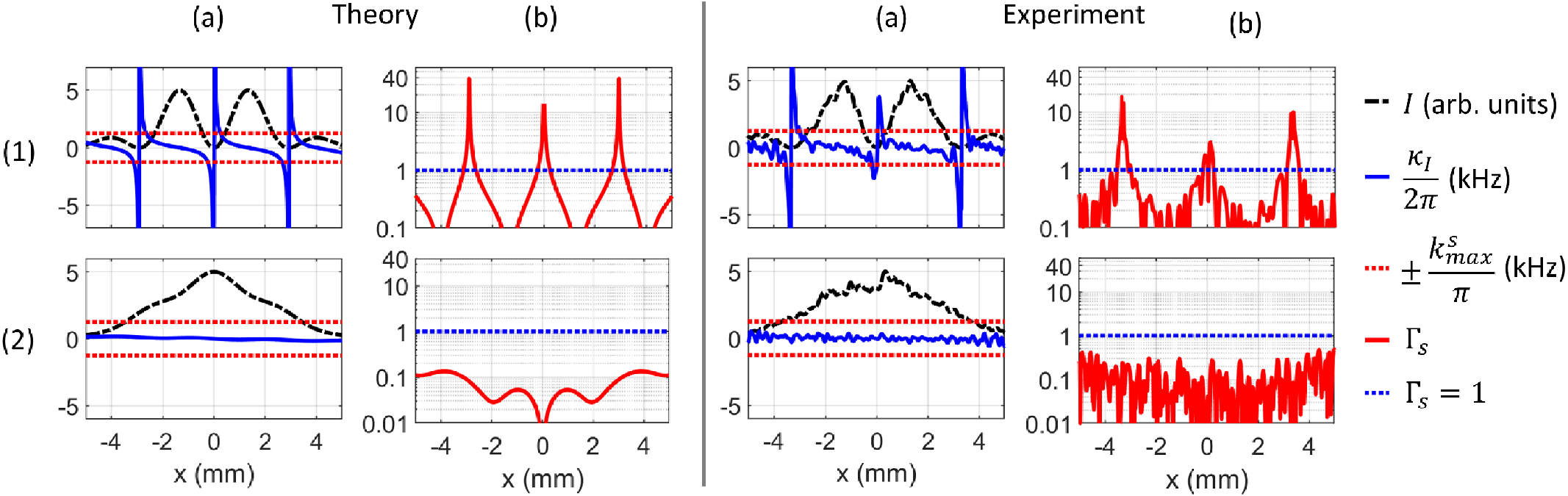}
\caption{\label{fig3}  Simulations [two left columns]  and experimental construction [two right columns] of  optical fields with [row (1)] and without [row (2)] supergrowth. 
Columns (a) show  x-line-cut of $I$ and $\kappa_I$, along with $2 k^s_{\max}$, identifying the supergrowing regions.   
Columns (b) show the x-line-cut of $\Gamma_s$  from Eq.~\eqref{Gamma_s_intensity_eq_sys}. 
In (b), the doted blue line separates the supergrowing region ($\Gamma_s(x)\geq1$) and non-supergrowing region ($\Gamma_s(x) < 1$). 
We are able to achieve a maximum supergrowth of $\Gamma_s\approx18.3$  in the experiment.
These plots demonstrate our ability to generate and control supergrowth in an experiment.   }
\end{figure*}

\emph{Experiment.}
---The schematic of the optical setup used to generate supergrowing optical fields is shown in Fig.~\ref{fig2}. An intensity-stabilized $795~\mathrm{nm}$ linearly polarized collimated laser beam (Toptica TA 100) passes through a beam expander using lenses $L_1$ and $L_2$. 
This beam is transferred to a computer-generated phase-only pixelated hologram (CGPPH) generated using a liquid-crystal-based SLM (Hamamatsu LCOS). 
A beam expander using the lenses $L_1$ and $L_2$ ensures the laser beam overfills the active area of the SLM. As a result, the CGPPH sees less spatial variation of intensity of the illuminating laser and a nearly uniform phase.  
To stabilize the laser intensity, an acousto-optic modulator with a PID controller is used, and polarization optics and attenuators are used to control the polarization and intensity of the laser beam (not shown in the schematic). 
The diffraction from the CGPPH is Fourier processed using a classical $4f$ processor consisting of lenses $L_3$ and $L_4$, and a precision pinhole $P$ (Thorlabs P1000K). 
The intensity distribution of the complex field at the image plane ($Z_I$) is measured using a CCD camera.
We have ensured the output intensity lies within the linear response of the detector.  

Phase and amplitude of the optical field at co-ordinates $(u,v)$ in the transverse plane $Z_F$ determine the  phase and amplitude of image plane ($Z_I$) Fourier components at $f_x=\frac{u}{\lambda f}$ and $f_y=\frac{v}{\lambda f}$ \cite{goodmanIntroductionFourierOptics2017}. 
The maximum spatial frequency at the image plane is limited by the diameter $d$ of the precision pinhole at $Z_F$, and the system bandlimit is $k^s_{\max}=\frac{\pi d}{\lambda f}$.

\emph{Generation of supergrowing functions.}---
The CGPPH used in the experiment for generating a supergrowing function uses encoding techniques described in \cite{Arrizon:07}. The function to be prepared in the experiment must satisfy two criteria. 
\begin{enumerate}
    \item Bandlimit: $k^f_{\max}<k^s_{\max}$. 
    \item Spatial-limit: The distribution of supergrowing field at the image plane $Z_I$ must remain within the image of SLM active area at $Z_I$.  
\end{enumerate}
It is a challenge to satisfy both of these criteria as limitations imposed in the spatial domain demand a scaling up in the frequency domain.  
Larger values of $k^f_{\max}$ are necessary to ensure maximum supergrowth $\Gamma_s\geq 1$ within the spatial limit.
Theoretically, $k^f_{\max}=k^s_{\max}$ is possible, but due to experimental imperfection and uncertainties, one cannot achieve this upper limit for a field   in experiments. 

We explore two different ways of constructing candidate supergrowing functions in a circularly symmetric geometry: 1)~heuristic analytical method and 2)~numerical simulations. 
In the first approach, we  start with a simple circularly symmetric analytical function with supergrowing features $f(\rho)=f_1(\rho)f_2(\rho)$, where $\rho^2=x^2+y^2$. 
Here,
$f_1(\rho)=\exp(\frac{-\rho^2}{2c^2})$, where $c$ is a constant, ensures a limited spatial area extent of $f(\rho)$; and $f_2(\rho)=\cos(k_{f_2}[\rho-\rho_0])+ i a \sin(k_{f_2}[\rho-\rho_0])$, where the shift $\rho_0$ is such that the maximum or minimum of $|f_2(\rho)|$ is at $\rho=0$. Also, $a$ is a parameter much bigger than 1.
The inspiration behind $f_2(\rho)$ comes from the $1D$ function $f(x)=(\cos x+i a\sin x)^N$, whose supergrowing and imaging properties have been studied in depth already \cite{jordanSuperresolutionUsingSupergrowth2020,karmakar2023supergrowth}.  
We choose $k_{f_2}\gg k_{f_1}$, where $k_{f_1}$ is four times the full-width-half-maximum of the Fourier transform of $f_1(\rho)$. The bandlimit criterion is satisfied as  
$k_{f_2}+k_{f_1}=b k^s_{\max}$, where $0.75\leq b\leq 0.95$. 
Then, $f(\rho)$ supergrows locally within the spatial limit when $a\gg1$. 

The ability of this analytical function to generate supergrowth is limited. 
Numerical simulations help us overcome this obstacle.  To find the appropriate fields that supergrow within the frequency and space limit on the image plane $Z_I$, we perform simulations using the Zernike polynomial-Bessel function basis for circular symmetric functions, and 
  simultaneously adopt the Helmholtz equation-based definition of local wavenumber/growth rate (Appendices E and F of Ref.~\cite{karmakar2023superoscillation}). A multi-objective cost optimization is performed to ensure 1) optimum extent of supergrowing 
  region within the image plane, 2) optimization of the ratio of maximum intensity within the supergrowing region and the maximum intensity in the image plane. The second condition ensures the effect of enhanced sidelobes, a limitation observed in superoscillation-based superresolution imaging, is curtailed \cite{rogersOptimisingSuperoscillatorySpots2018}.

 \begin{figure*}[t]
\centering
\includegraphics[width=\textwidth]{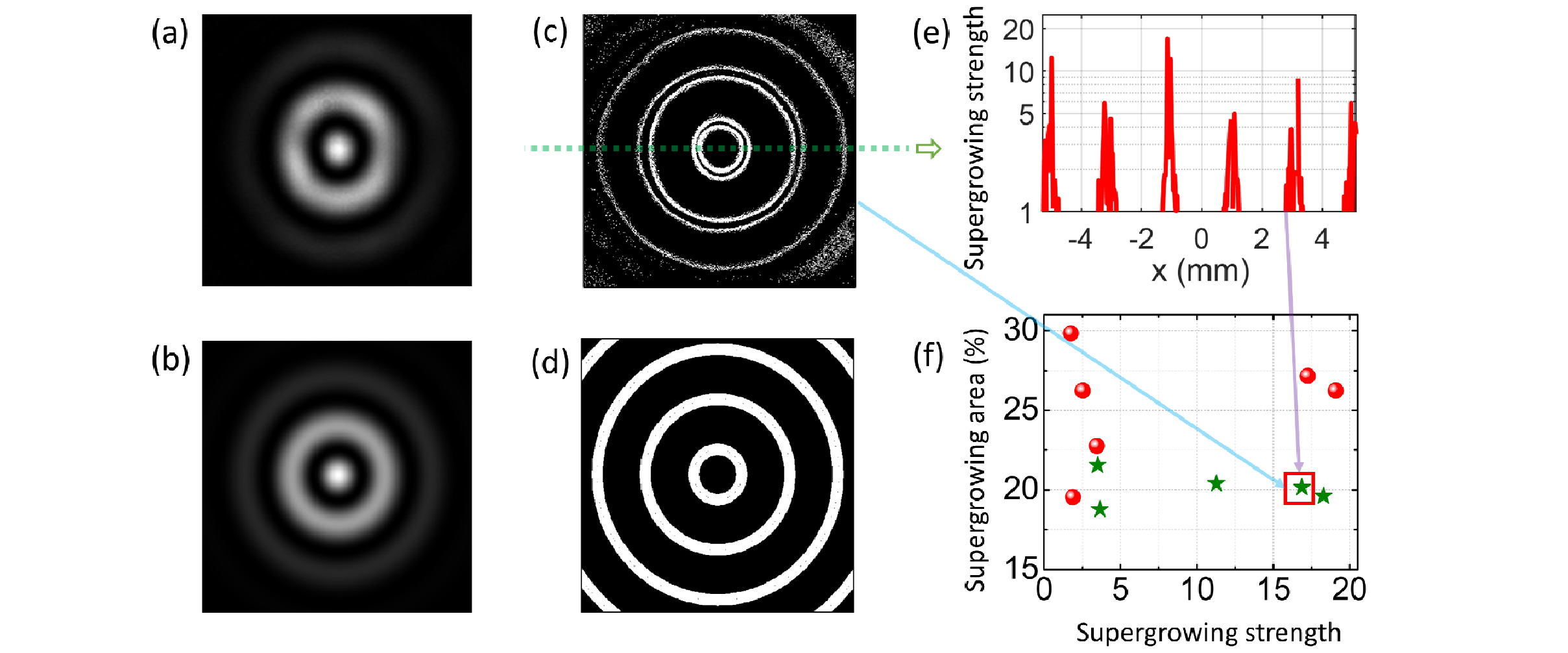}
\caption{\label{fig4} Intensity of a synthesized  field at the image plane $z_I$ (Fig.~\ref{fig2}), measured using the CCD, is shown in (a) and the corresponding analytical function is shown in (b). 
Panels (c) and (d) identify the supergrowing regions of measured and theoretical fields, respectively, in white.
Supergrowing strength $\Gamma_s$ of supergrowing regions identified in (c) along a horizontal line cut (highlighted using a semi-transparent dashed green line) is shown in (e). We have averaged four adjacent rows of pixels to get this line cut to reduce the effect of noise. Panel
(f) shows the quality of supergrowth of multiple fields, synthesized in the experiment, represented here by plotting supergrowing area $\Upsilon_s$ and maximum supergrowing strength $\Gamma_s$ along the x-linecut of each field. 
Here Red (green) spheres (stars) represent the function obtained using the simulations (heuristic method). 
The field shown in (a)-(e) is highlighted using a red square in (f). 
Simulations provide greater control over the supergrowing area of target fields, and corresponding experimental synthesis shows $\Upsilon_s$ between $19.5\%$ and $29.8\%$.   
}
\end{figure*}
 
\emph{Supergrowing area $\Upsilon_s$ of measured image.}--- Supergrowing area denotes the fraction of the area of a complex field that is supergrowing at the image plane. $\Upsilon_s$ is another tool for quantifying the quality of the supergrowing field--- a higher value of $\Upsilon_s$ is better. But this definition of $\Upsilon_s$ is inadequate for images obtained from the experiment due to noise. Let $f(x,y)$ be the complex field at the image plane (Fig.~\ref{fig2}).  In the experiment, the CCD measures the spatially discretized and noisy version of intensity $I(x,y)=|f(x,y)|^2$ as $I_0(m,n)$, where $m,~n$ are matrix indices of the matrix representing CCD pixels. We have ensured the maximum spatial frequency of the field is much smaller than the frequency of the CCD pixels so that the Nyquist criterion is not violated. One can express $I_0(m,n)=\eta_d(m,n)I(m,n)+I_\xi(m,n),$
where, $\eta_d(m,n)$ is the detection efficiency of the pixel $(m,n)$, $I_\xi(m,n)$ is the noise, and $I(m,m)$ is discretized $I(x,y)$. 
The intensity quantization at each pixel is ignored in this representation. 

Let the maximum value of the noise $I_\xi(m,n)$ be $I_{\xi0}$. One cannot consider pixels with intensities $I_0(m,n)\leq I_{\xi0}$. The supergrowing area of measured images can be defined as 

\begin{equation}
    \Upsilon_s=\frac{ \zeta_s}{ \zeta},
    \label{Eq:SG_area_exprmnt}
\end{equation} 
where $\zeta$ is the total number of pixels with $I_0(m,n)\geq I_{\xi0}$, and $\zeta_s$ is the total number of pixels with  $\Gamma_s(m,n)\geq1$ and $I_0(m,n)\geq I_{\xi0}$.

\emph{Results.}--- We have prepared multiple bandlimited complex functions with finite spatial extent using analytical expression and simulation. 
The fields realized in the experiment have a diverse range of supergrowing strength $\Gamma_s$ and area $\Upsilon_s$. Figure~\ref{fig3} demonstrates the ability to control the supergrowth of the synthesized field (with respect to $k^s_{\max}$) in the experiment.  
Row (1) of the figure shows a field with maximum $\Gamma_s\approx40$($\approx18.3$, in the experiment), while the field shown in row (2) is not supergrowing ($\Gamma_s<1$).

A transverse distribution of intensity and corresponding supergrowing regions at the image plane $Z_I$ [Fig.~\ref{fig2}] of a representative complex field (both in theory and experiment) are demonstrated in Figs.~\ref{fig4}(a)-(d).
The supergrowing area of this field is  $\Upsilon_s\approx23.6\%$ ($\approx20.16$, in the experiment). The experiment exhibits lower $\Upsilon_s$ and double-ring patterns [Fig.~\ref{fig4}(c)] because pixels with intensities  less than the cutoff $I_{\xi0}$ [Eq.~(\ref{Eq:SG_area_exprmnt})] are ignored.
Note, for such $2D$ images of intensity $I(x,y)$, we calculate the local intensity growth rate  as $\kappa_I(x,y)=|\nabla\log\big[I(x,y)\big]|$. 
By definition,  supergrowth is more sensitive to fluctuations at low intensities. 
In the experiment, noise/distortions present at intensities comparable to the noise floor can lead to identifying more supergrowing regions than the analytical field.
The definition of $\Upsilon_s$ [Eq.~(\ref{Eq:SG_area_exprmnt})] reduces this effect by neglecting low-intensity pixels. 





Measured values of the supergrowing area $\Upsilon_s$ and the maximum values of supergrowing strength $\Gamma_s$ for multiple supergrowing fields synthesized in the experiment are shown in Fig.~\ref{fig4}(f). 
One can observe that the functions obtained from the simulations, in general, can attain larger values of $\Upsilon_s$. This is consistent with the multiobjective optimization conditions described previously.
We have achieved a significantly larger supergrowing area (up to $\Upsilon_s\approx 29.8 \%$) as well as a maximum local growth rate reaching up to $\approx19.1$ times of the  bandlimit of the system. 
A supergrowing strength of $\sim19$ is highly desirable since optical spots with high local growth rates help us image subwavelength objects.

\emph{Discussion.}---
We have demonstrated the experimental synthesis of supergrowing optical fields. 
We also prescribe comprehensive methods and parameters to measure and characterize the supergrowth in an experiment.
In this context, we have clarified the distinction between the supergrowth of an optical field and the concept of supergrowth in a diffraction-limited optical system. 
While the latter provides a stronger constraint on the condition of supergrowth, we are still able to achieve  local growth rates more than an order of magnitude higher in our setup compared to the system bandlimit. 
Additionally, our setup shows excellent control over supergrowth since we realize optical fields with a diverse range of supergrowing areas and strengths.
Supergrowing regions inherently can contain exponentially higher intensities compared to superoscillating regions \cite{jordanSuperresolutionUsingSupergrowth2020,karmakar2023supergrowth}. 
Therefore, supergrowth will alleviate the effect of enhanced sidelobes plaguing superoscillation-based imaging and improve SNR in experiments. 
This letter lays the groundwork for generating supergrowth in the lab and serves as the foundation for a novel superresolution imaging system.

A natural next step is to realize supergrowth-based subwavelength object reconstruction proposed in Ref.~\cite{karmakar2023supergrowth}.
Also, supergrowth being a weak value amplification of momentum \cite{jordanSuperphenomenaArbitraryQuantum2022}, generating optical fields with higher supergrowing strengths could be useful in a multitude of applications as well, including communication and quantum technologies. 
Furthermore, an optical field spot with simultaneously optimized supergrowing and superoscillating properties could prove to be extremely beneficial for superresolution imaging. 

We acknowledge insights provided by Abhishek Chakraborty, Arunabh Mukherjee, and Valeria Viteri-Pflucker in this project. This work was supported by the AFOSR grant \#FA9550-21-1-0322 and the Bill Hannon Foundation.

\bibliography{refs}
\end{document}